\def\Eq#1{Eq.~(\ref{#1})}
\def\Tr{\rm Tr}
\def\pmu{p_{\mu}}
\def\gammamu{\gamma_{\mu}}
\def\gammanu{\gamma_{\nu}}
\def\nhati{\hat{n}^{i}}
\def\nhatik{\hat{n}^{i_{k}}}
\def\sumdot{\sum_{i_{N}=1}^{2D}\cdots\sum_{i_{1}=1}^{2D}}
\def\spinfactor{(1-i\gamma . \hat{n}^{i_{N}})\cdots
(1-i\gamma . \hat{n}^{i_{1}})}
\def\sumk{\sum _{k=1}^{N}}
\def\Gtilde{\tilde{G}}
\def\Dp{\frac{d^{D}p}{(2\pi)^{D}}}
\def\limNinf{\lim_{N\rightarrow\infty}}
\def\normFactor{{\Bigl(\frac{1}{2D}\Bigr)}^{N}}
\def\limAzero{\lim_{a\rightarrow0}}
\def\gaugefactor{(U_{\hat{n}^{N}} \cdots U_{\hat{n}^{1}})}
\def\smu{\sin (\pmu a)}
\def\DeltaN{\Delta _{n}}
\def\GtildeN{\tilde{ G_{n} } }
\def\nhatimu{\nhati _{\mu}}
\def\nhatikmu{\nhatik _{\mu}}
\def\LatPint{ \int_{\frac{-\pi}{a}}^{\frac{\pi}{a}} }
\def\Kbarh{\bar{K}^{h}}
\def\kmu{k_{\mu}}
\def\Kbarhmu{\Kbarh_{\mu}}
\def\Wterm{\sum_{\mu = 1}^{D} ( 1 - \cos (\pmu a))}
\documentstyle[12pt]{article}
\input{epsf}

\begin{document}
\def\thepage{}
\vbox{\hfill hep-lat/9701010}
\begin{center}
{\LARGE{Random Walk Representation of the Lattice Fermionic 
Propagators and the Quark Model}}\\
\medskip
\medskip
\large{{Vikram Vyas} \\
\medskip
The Ajit Foundation\\
Jaipur 302 018, India\\
{\it e-mail: ajit@unv.ernet.in}}
\end{center}
\begin{abstract}
A representation of the  continuum 
fermionic propagator as a sum of directed random walks on a lattice is presented.
Also a random walk representation for the lattice fermionic propagators is 
developed for the case of the naive, the Wilson, and the 
Kogut-Susskind fermions.  
For the naive fermions the phenomenon of 
fermion doubling appears as having $2^{D}$ distinct spin factors being  
associated with a single path in $D$-dimensions. In the case of the 
Wilson and the Kogut-Susskind fermions, in the naive continuum limit, the 
path integral representation coincides with the path integral 
representation for the continuum fermionic propagator. Using this 
representation the Green's functions of lattice QCD involving quark operators are 
written as a sum over the paths of valence quark, the gauge fields and 
the sea quarks being integrated out. Possible advantages of such a representation 
are illustrated by showing how one can use numerical simulations to 
obtain a heuristic insight into the relationship between QCD and the 
constituent quark model.
\end{abstract}

\newpage
\renewcommand{\thepage}{\arabic{page}}
\setcounter{page}{1}

\section{Introduction}

It seems increasingly likely that in the near future the lattice
simulations will provide 
an accurate numerical solution of Quantum Chromodynamics (QCD). 
Complimenting these numerical investigations are many attempts at 
obtaining a heuristic understanding of the dynamics of QCD by trying 
to isolate those fluctuations in the gauge fields that may be 
responsible for the confinement and chiral symmetry breaking~\cite{Teraflop}.
A similar interplay between numerical simulations and heuristic 
understanding is lacking for the problems involving 
the quark degrees of freedom. Thus there are many striking features 
of strong interactions, an incomplete list would include the apparent 
successes of the non-relativistic quark model and the 
Okubo-Zweig-Iizuka rule~\cite{sModel}, for which the 
numerical simulations provide little or no heuristic guidance.

The reasons for this are familiar, the fermionic degrees of freedom being 
elements of Grassmann algebra cannot be simulated numerically. These 
degrees of freedom have to be integrated out from the functional 
integral before it can be numerically evaluated. The result of 
the integration is expressed in terms of the fermionic determinant 
and the fermionic propagator~\cite{Creutz, CJR, KogutRMP, book1, book2}. But the 
numerical calculation of the determinant and the inverse of a 
fermionic matrix  seems to offer little scope 
for visualizing the  degrees of freedom that gave rise to 
them in the first place. 

In this paper I suggest an alternate way of dealing with the fermionic 
propagator which is more amenable to  heuristic investigations. 
The approach is based on the path integral representation, or the 
equivalent random walk representation\footnote{In this paper I will 
interchangeably use the words path-integral and random walk 
representation.}, of the
fermionic propagator~\cite{Poly, TJ, Ambjorn, Report}. One way of 
obtaining a random walk representation for the lattice fermions is to 
expand the propagator for the Wilson fermions in a Neumann 
series~\cite{ Hasenfratz81, Kuti82, KogutRMP}. An alternate approach, 
which is more general, is based on Polyakov's observation~\cite{Poly}
that the path integral for a fermion can be regarded as a natural generalization of the 
path integral for a scalar particle by including an appropriate 
spin factor. In this paper I will take the latter approach.

With in such an approach it is natural to ask the question that how 
does the phenomenon of fermion doubling on a lattice~\cite{KSmit, NN} manifests 
itself in the language of  path integrals. To answer this, in Sec.~2, 
first I represent the  continuum fermionic propagator as a sum over 
directed random walks on a lattice. Then in Sec.~3 and~4,
 I consider  the path integral representation for the lattice fermionic 
propagators, this includes the so called naive fermions, the Wilson 
fermions, and the Kogut-Susskind fermions. In the case of the naive fermions
the expected phenomenon of fermion  doubling appears as the 
existence of  $2^{D}$ distinct spin-factors for a single path in 
a $D$-dimensional hypercubic lattice. Continuing the analysis in 
Sec.~4, it is shown that for
the case of the Wilson and the Kogut-Susskind fermions, in the naive 
continuum limit, only one of the $2^{D}$ spin-factor survives and the 
path integral representation of the propagators coincides with that of 
the continuum fermionic propagator developed in Sec.~2.

Having developed a path integral representation for a free fermionic 
propagator it is easy to extend it to the case when a gauge field is 
defined on a lattice~\cite{Wilson74, feynLH}.
Using the resulting representation of the fermionic propagator, in 
Sec.~5, I write the Green's functions involving quark fields as a sum over 
the paths of the valence quarks. This way of writing the Green's 
functions in which the integration over the gauge field is carried
out before summing over the quark paths is a close transcription, 
which allows for numerical explorations, of the heuristic way in 
which Wilson~\cite{Wilson74} motivated his criterion for the 
quark confinement. Perhaps the advantage of this approach is that it allows 
us to probe the theory using a language, the paths of the valence 
quarks, which is easy to visualize. This is illustrated in Sec.~5 by 
indicating how the numerical simulation can provide a insight into 
the relationship between QCD and the constituent quark model.
I state my conclusions in the last section.

\section{Representing Continuum Fermionic Propagator as Random Walks 
on a Lattice}

The fermionic propagator can be represented as a path integral either 
using an appropriate spin-factor~\cite{Poly}, or equivalently as a sum over 
directed random walks~\cite{TJ, Ambjorn, Report}.  In this 
section I will represent the continuum Euclidean Fermionic 
propagator as a sum over directed random walks on a $D$ dimensional hypercubic
lattice. Following~\cite{Report} I start with the 
Fermionic propagator in the momentum space 
\begin{equation}
\Delta (p, m) = \frac{1}{ m + p_{\mu}  \gamma_{\mu}},
\label{contProp}
\end{equation}
where $\gammamu$ are the Euclidean Dirac matrices satisfying the 
anticommutation relation
\begin{equation}
\{ \gammamu, \gammanu \} = -2 \delta_{\mu \nu}.
\label{dAlg}
\end{equation}
The propagator is now written as a Laplace transform
\begin{equation}
\Delta (p,m) = \frac{1}{D} \int_{0}^{\infty} \exp{ \{-\frac{mL}{D}\} } 
\tilde{G} (L, p)dL,
\label{LTcontProp}
\end{equation}
\begin{equation}
\tilde{G}(L,p) = \exp{ \{ \frac{-\gammamu \pmu L}{D} \} },
\label{Gtilde}
\end{equation}
where $D$ is the dimension of the Euclidean space which, unless stated 
otherwise, will be taken to be four.
The motivation for expressing the propagator as a Laplace transform 
is that the Fourier transform of \Eq{Gtilde}
can be then interpreted as the probability amplitude for the particle to 
reach the point $x$ by traveling  along paths of length $L$. 
With this in mind, the Laplace transform is 
expressed as
\begin{equation}
\tilde{G}(L,p) = \lim _{N \rightarrow 
\infty}\left(1-\frac{\gammamu \pmu L}{DN}\right)^{N}. 
\label{prRepLT}
\end{equation}
The individual terms appearing in \Eq{prRepLT} can be written as
\begin{equation}
	1 - \frac {\gamma . p L}{ND} = \frac {1}{2D}\sum_{i=1}^{2D}
	(1 - i \gamma . \nhati) (1 - i \nhati . p \frac{L}{N}),
	\label{indLattice}
\end{equation}
where $\nhati$ are the directions available to a particle on a 
$D$-dimensional hypercubic Euclidean lattice, and for $D=4$ 
can be taken to be:
\begin{equation}
\begin{array}{lclccr}
	 \hat{n}^{1}&=&(1,0,0,0),\quad\hat{n}^{5}&=&(-1,0,0,0),\\
	 \hat{n}^{2}&=&(0,1,0,0),\quad\hat{n}^{6}&=&(0,-1,0,0),\\
	 \hat{n}^{3}&=&(0,0,1,0),\quad\hat{n}^{7}&=&(0,0,-1,0),\\
	 \hat{n}^{4}&=&(0,0,0,1),\quad\hat{n}^{8}&=&(0,0,0,-1).\\
	 \end{array}
	 \label{nhat}
\end{equation}
The identity, \Eq{indLattice}, can be verified by using the 
above choice  of $\nhati$. 

This should be contrasted with the following \cite{Report} representation
\begin{equation}
1 - \frac {\gammamu \pmu L}{ND} = \int d\hat{n}
                                 (1-i\gamma.\hat{n})
                                 (1-i\hat{n}.p \frac{L}{N})
\label{cRWalk}
\end{equation}
where the integration is over all the directions in $D$-dimensional 
Euclidean space and results in a representation of the continuum 
fermionic propagator as a sum over paths  in the continuum.

Substituting \Eq{indLattice} into \Eq{prRepLT} leads to
\begin{equation}
	\tilde{G}(L,p) = \prod_{k=1}^{N}\left\{ \frac {1}{2D}\sum_{i=1}^{2D}
	(1 - i \gamma . \nhatik) (1 - i \nhatik . p \frac{L}{N})\right\},
	\label{Rwalk1}
\end{equation}
which in the limit $N\rightarrow\infty$, for a fixed $L$, can be 
writen as
\begin{eqnarray}
	 \tilde{G}(L,p) &=& 
	 {(\frac{1}{2D})}^{N}\sumdot\{\spinfactor \nonumber\\ 
	 & &\times\exp{(-i\frac{L}{N} p.\sum_{k=1}^{N}\nhatik)}\}. 
	\label{Rwalk2}
\end{eqnarray}
Taking the Fourier transform of \Eq{Rwalk2} leads to
\begin{eqnarray}
	\Gtilde (L, x-y) &=& \limNinf \normFactor \sumdot \{\spinfactor\} 
	\nonumber\\
	    &&\times \delta ^{D}(x-y -\frac{L}{N}\sum_{k=1}^{N}\nhatik).               
	\label{FTRwalk}
\end{eqnarray}
This, as anticipated, can be interpreted as a sum over paths of 
length $L$ starting at the point $y$ and ending at the point $x$. The 
paths in \Eq{FTRwalk} are defined on a $D$ dimensional Euclidean 
lattice, with each step in the direction $\nhati$ there is a 
spin-factor $(1 - i \gamma . \nhati)$ associated with it. The step 
size defines the lattice constant
\begin{equation}
	a = \frac{L}{N}.
	\label{defA}
\end{equation}

In view of \Eq{defA} the fixed length propagator can be 
written as
\begin{eqnarray}
	\Gtilde (L, x-y) & = & \Gtilde (N;a;x-y), \nonumber  \\
           \Gtilde (N;a;x-y) & = & \limAzero \normFactor \sumdot \spinfactor 
	                 \nonumber\\
	                 &&\times \frac{\delta ^{D}(x-y, 
	                 a\sum_{k=1}^{N}\nhatik)}{a^{D}}.
	\label{latGtilde}
\end{eqnarray}
In \Eq{latGtilde} the Dirac-$\delta$ function has been replaced with 
the Kronecker-$\delta$ function.

Using \Eq{LTcontProp} and replacing the integration over $L$ by a summation 
over $N$ leads to the following expression for the continuum fermionic 
propagator in the coordinate space
\begin{equation}
	\Delta (x-y, m) = \limAzero \frac{a}{D} \sum_{N=1}^{\infty}
	                  \exp(-\frac{maN}{D})\Gtilde (N;a;x-y).
	\label{latRepCprop}
\end{equation}
Though the sum in the above equation 
starts from $N=1$ but in the limit $a \rightarrow 0$, which 
is equivalent to the $\parallel x-y \parallel >> a$, 
only terms with very large value of $N$ contributes because of the 
presence the $\delta$-function in \Eq{latGtilde}.
Above representation of the continuum fermionic propagator as a sum 
over directed random walks on a lattice can be extended to the case when a 
gauge field is defined on a lattice. Following~\cite{Wilson74}, see 
also~\cite{feynLH}, 
the continuum fermionic propagator in 
the presence of a background gauge field, $U$, which is defined on a 
lattice can be written as:
\begin{equation}
	\Delta (x-y;m;U) = \limAzero \frac{a}{D} \sum_{N=1}^{\infty}
	                   \exp (-\frac{maN}{D})\Gtilde (N; a; x-y; U),
	\label{CferLatG}
\end{equation} 
where $\Gtilde(N; a; x-y; U)$ is given by
\begin{eqnarray}	
	\Gtilde (N; a; x-y; U) &=& \normFactor \sumdot \spinfactor 
	\nonumber \\
	&&\times \gaugefactor \frac{\delta^{D}(x-y, 
	a\sum_{k=1}^{N}\nhatik ) }{a^{D}} 
	\label{LGGtilde}
\end{eqnarray}
where $U_{\hat{n}^{k}}$ is a $SU(N)$ matrix associated with the link 
of the lattice that corresponds to the $k_{th}$ step of the random 
walk.

It is interesting to note that in the quenched approximation, where 
one neglects the contribution of the fermionic determinant, using 
above representation of the fermion propagator one can give a 
non-perturbative definition of QCD that keeps the fermionic degrees 
of freedom in continuum while defining the gauge degrees of freedom on 
a lattice. This would be of significance if one could also define the 
fermionic determinant in a similar manner (for some attempts along 
this direction see~\cite{Lunev}), for this would give a 
non-perturbative definition of QCD that avoids the problem of fermion 
doubling.

\section{Random Walk Representation of the Naive Lattice Fermionic 
Propagator}

In this and the next section a representation of the lattice fermionic 
propagators as a sum over paths on a lattice will be obtained. 
The naive lattice fermionic 
propagator~\cite{Creutz, book1, book2} in 
the momentum space is given by 
\begin{equation}
	\DeltaN (p, m) = \frac {1} {m + \frac {1}{a}{\gammamu\smu} },
     \label{NProp}
\end{equation}
where $a$ is the lattice constant and the $\gammamu$s are the 
Euclidean Dirac matrices satisfying \Eq{dAlg}. It is well known that
in the continuum limit \Eq{NProp} describes the propagation of $2^{D}$ distinct 
fermions~\cite{KSmit}, an example of the general phenomenon of 
fermion doubling~\cite{NN}. Writing the naive fermionic propagator 
as a sum over paths will allow us to understand this phenomenon
in the language of path integrals.

The development of the path integral representation for the naive 
lattice fermionic propagator starts, as before, with a formal Laplace transform of \Eq{NProp}
\begin{eqnarray}
	\DeltaN (p, m) & = & \frac{1}{D} \int_{0}^{\infty} \exp{ \{-\frac{mL}{D}\} } 
                        \GtildeN (L, p)dL, \label{NLT} \\
	\GtildeN (L, p) & = & \exp{ \{ \frac{-\gammamu \smu L}{Da} \} }
	\label{GtildeN}
\end{eqnarray}
and with the identity
\begin{equation}
	1 - \frac {\gammamu \smu L}{DNa} = \frac {1}{2D}\sum_{i=1}^{2D}
	(1 - i \gamma . \nhati) (1 - i \nhatimu \smu  \frac{L}{Na}),
	\label{NLattice}
\end{equation}
where $\nhati$ are the unit vectors, along both the positive and 
negative directions of a $D$-dimensional Euclidean hypercubic lattice 
and are defined in \Eq{nhat}. Using \Eq{GtildeN} and \Eq{NLattice}, as in section 2, 
we can write $\GtildeN(L, p)$ as
\begin{eqnarray}
	\GtildeN (L, p) &=& \limNinf {(\frac{1}{2D})}^{N}\sumdot\{\spinfactor \nonumber\\ 
	 & &\times\exp{(-i\frac{L}{Na} \smu . \sumk \nhatikmu)}\}.
	\label{RwalkGN}
\end{eqnarray}
The Fourier transform of \Eq{RwalkGN}, defined by
\begin{equation}
	\GtildeN (L, x-y) = \LatPint \Dp \exp(ip.(x-y))\GtildeN (L, p)
	\label{latFT}
\end{equation}
with $x$ and $y$ belonging to the lattice, can be written as
\begin{eqnarray}
	\GtildeN (L, x-y) & = & \limNinf \normFactor \sumdot \spinfactor 
	\nonumber \\
	                  &  &\times I[x-y, \{ \nhati \} ], 
	\label{FTGN}
\end{eqnarray}
where $I[x-y, \{\nhati\}]$ is given by 
\begin{equation}
	I[x-y, \{ \nhati \} ] = \LatPint \Dp \exp (i p. (x-y))\exp\{-i\frac{L}{Na} 
	\smu . \sumk \nhatikmu \}.
	\label{defI}
\end{equation}
Next  consider $\GtildeN(L, x-y)$
 in the limit $a\rightarrow 0$. For this it will be convenient to, 
using the periodicity of the lattice momentum 
space, shift the  range of integration from $\frac{-\pi}{a} \leq 
\pmu < \frac{\pi}{a}$ to $\frac{-\pi}{2a}\leq \pmu < \frac{3\pi}{2a}$. 
Also the range of for each $\pmu$ is divided into two 
regions, in the first region $\pmu$ ranges from $\frac{-\pi}{2a} 
\leq \pmu < \frac{\pi}{2a}$, and in the second region it ranges 
from $\frac{\pi}{2a} \leq \pmu < \frac{3\pi}{2a}$. In this manner the 
lattice momentum space divides into $2^{D}$ hypercubes centered 
around points for which $\pmu$, the $\mu$th component, is either 
$zero$ or $\frac{\pi}{a}$. These centers will be denoted by the 
vector $\Kbarh$. Using this division of the momentum space \Eq{defI} 
can be written as
\begin{equation}
	I[x-y; \{ \nhati \}] = \sum _{h=1}^{2^{D}} I_{h}[x-y; \{ \nhati \}],
	\label{divideI}
\end{equation}
with $I_{h}[x-y; \{ \nhati \}]$ given by
\begin{equation}
	I_{h}[x-y; \{ \nhati \}] = \int_{h} \Dp \exp ( -i\frac{L}{Na} \smu 
	\sumk \nhatikmu ) \exp (ip.(x-y)).
	\label{defIh}
\end{equation}
In \Eq{defIh}, as noted above, the $\mu_{th}$ component of $\pmu$ 
lies either in the range $\frac{-\pi}{2a} 
\leq \pmu < \frac{\pi}{2a}$ or in the range $\frac{\pi}{2a} \leq 
\pmu < \frac{3\pi}{2a}$.
 Defining a new integration variable $\kmu$,
\begin{equation}
	\pmu = \Kbarhmu  + \kmu,
	\label{defkmuh}
\end{equation}
allows one to write \Eq{defIh} as
\begin{equation}
	I_{h} = e^{i\Kbarh . (x-y)}\int_{\frac{-\pi}{2a}}^{\frac{\pi}{2a}} \frac{d^{D}k}{(2\pi)^{D}}
	           \exp ( -i \frac{L}{N} \frac{\sin (\Kbarhmu a + \kmu 
	           a)}{a} \sumk \nhatikmu ) \exp ( ik.(x-y) ).
	\label{syIh}
\end{equation}
Taking the limit $a \rightarrow 0$ of the above expression leads to
\begin{equation}
	\limAzero I_{h}[x-y]; \{ \nhati \}] = e^{i\Kbarh . (x-y) }
	                            \delta^{D}( (x-y) - 
	                            \frac{L}{N}\cos(\Kbarhmu a) \kmu . 
	                            \sumk \nhatikmu ),
	\label{limazeroIh}
\end{equation}
where $\cos(\Kbarhmu a)$ is either $+1$ or $-1$, depending on 
whether $\Kbarhmu$ is $0$ or $\frac{\pi}{a}$.  Thus $\GtildeN(L, x)$ 
in the limit $a\rightarrow0$ has the following form
\begin{eqnarray}
	\GtildeN (L, x) & = & \normFactor \sumdot \spinfactor 
	\nonumber \\
	                  &  &\times \sum_{h=1}^{2^{D}}
	                \{ e^{i\Kbarh . (x-y) }
	                           \delta^{D}( x - 
	                            \frac{L}{N}\cos(\Kbarhmu a) \kmu . 
	                            \sumk \nhatikmu ) \}.
	\label{almostRW} 
\end{eqnarray}
$\GtildeN(L, x)$ can now be interpreted as 
a sum over $2^{D}$ 
distinct random walks of length $L$  but for the factor of 
$\cos(\Kbarhmu)$. This factor can removed from the $\delta$ function 
by suitable redefinition of the spin-factors. To see this consider 
the case when the $\nu_{th}$ component of $\Kbarh$ is 
$\frac{\pi}{a}$, then substitute the set of unit vectors $\{\nhati\}$ 
by a new set of vectors obtained by changing the sign of the 
$\nu_{th}$ component of the vectors $\{\nhati\}$. This transformation 
keeps \Eq{almostRW} invariant as it corresponds merely to relabeling 
of the vectors  $\{\nhati\}$. Because of this the 
spin factor for a link
$ 1 - i(\cdots + \gamma_{\nu}\nhati_{\nu} + \cdots)$, where no summation over 
$\nu$ is implied, changes to $1 - i(\cdots - \gamma_{\nu} \nhati_{\nu} + \cdots)$. 
This can be brought to its original form by doing a similarity 
transformation on the gamma matrices
\begin{equation}
	\gamma_{\nu} = -\gamma_{\nu}^{h} = S^{h}\gamma_{\nu} (S^{h})^{-1}.
	\label{defgammah}
\end{equation}
As a result of these transformations $\GtildeN(L, x)$ can be written as
\begin{eqnarray}
	\GtildeN (L, x) & = & \sum_{h = 1}^{2^{D}} \Gtilde_{h} (L, x)
	\label{fdGN},  \\
	\Gtilde_{h} (L, x) & = & \normFactor \sumdot 
	                                     (1 - i 
	                                     \gamma^{h}.\hat{n}^{i_{N}} )
	                                     \cdots (1 - i \gamma^{h} . 
	                                     \hat{n}^{i_{1}})
	\nonumber  \\
	                             &  & \times \delta^{D}(x - 
	                             \frac{L}{N}\sumk \nhatik),
	\label{GtildeH}
\end{eqnarray}
which can be interpreted as a sum over
paths of length $L$ with the step-size $a$ where $a$ is the 
lattice constant. Finally writing the formal Laplace transform, 
\Eq{NLT}, as a sum over lengths in lattice units leads to
\begin{eqnarray}
	\limAzero \Delta_{n}(x, m) & = & \frac{a}{D}\sum_{N=1}^{\infty}
	                                                    \exp(-\frac{maN}{D})
	                                                    \sum_{h = 1}^{2^{D}} 
	                                                    \Gtilde_{h} (N, x),	                                                    
	\nonumber  \\
	\Gtilde_{h} (N, x)& = & e^{i\Kbarh .x}\normFactor \sumdot 
	                                     (1 - i 
	                                     \gamma^{h}.\hat{n}^{i_{N}} )
	                                     \cdots (1 - i \gamma^{h} . 
	                                     \hat{n}^{i_{1}})
	\nonumber  \\
	                        &  &  \times \frac{\delta^{D}(x, a\sumk 
	                        \nhati)}{a^{D}}.
	\label{RWNprop}
\end{eqnarray}
The above equation shows that in the continuum limit the naive fermionic propagator  
represents propagation of $2^{D}$ distinct fermions. In the same 
limit the propagator of 
each of these fermions can be written as a sum over paths, or random 
walks, on the lattice. Different species differ by the spin-factor 
associated with them and by an overall phase factor. Since the phase 
factor can be absorbed in the normalization of the single particle 
wave function and the spin factors differ only by a 
similarity transformation of the $\gamma$-matrices, therefore we have 
in \Eq{RWNprop} a simultaneous propagation of $2^{D}$ Dirac particles. 
Further analysis along the line of Ref. \cite{KSmit} revels that these 
doublers appear in pairs with opposite chirality. 
\section{Random Walk Representation for Wilson 
and Kogut-Susskind Fermions}

Having elucidated the phenomenon of fermion doubling in the language 
of path integrals for fermions, it is natural to see how the Wilson and 
the Kogut-Susskind fermions avoid or mitigate this 
problem. I start with the Wilson fermions, the corresponding propagator in 
the momentum space is~\cite{Creutz, book1, book2}
\begin{equation}
	\Delta_{W} (p) = \frac{1}{m + \frac{1}{a}\smu  + 
	\frac{1}{a}\Wterm}.
	\label{Wprop}
\end{equation}
The development of a path integral representation for the Wilson 
fermions is similar to the previous two cases, again one writes the 
propagator as a formal Laplace transform
\begin{eqnarray}
	\Delta_{W}(p) & = & \frac{1}{D}\int_{0}^{\infty} dL 
	\exp(\frac{-mL}{D}) \Gtilde_{W} (L, p),
	\label{LTWprop}  \\
	\Gtilde_{W} & = & \exp(\frac{L}{Da}(\gammamu \smu + \Wterm).
	\label{DefGW}
\end{eqnarray}
Then using the following identity
\begin{eqnarray}
\lefteqn{1 - \frac{L}{DNa} (\gammamu \smu + \Wterm) = } 
\label{wSfactor}\\
& &\frac{1}{2D}\sum_{i = 1}^{2D}\{ ( ( 1 - i \gamma . \nhati ) 
( 1 - i\frac{L}{Na}\nhatimu \smu )) -\frac{L}{DNa}\Wterm\}, \nonumber 
\end{eqnarray}
which can be verified with $\nhati$ as the unit vectors defined by \Eq{nhat}, to write 
$\Gtilde_{W} (L, p)$ as
\begin{eqnarray}
	\Gtilde_{W} (L, p) & = & \limNinf \normFactor \sumdot \spinfactor 
	 \label{GW}\\
	 &  & \times \exp( -i \frac{L}{Na} \smu . \sumk \nhatikmu 
	 -\frac{L}{Da} \Wterm ).\nonumber
\end{eqnarray}
Taking the Fourier transform of  \Eq{GW} leads to
\begin{eqnarray}
	\Gtilde_{W}(L, x) & = &  \limNinf \normFactor \sumdot \spinfactor 
	\nonumber \\
	 &  & \times I_{W}[x, \{ \nhati \} ],
	\label{FTGW}  \\
	I_{W}[x, \{ \nhati \} ]& = & \LatPint \exp (i p.x) 
	                                       \exp( -i \frac{L}{Na} \smu . \sumk \nhatikmu \nonumber \\
	                                 & &  -\frac{L}{Da} \Wterm ).
	\label{IW}
\end{eqnarray}
\Eq{IW} differs from \Eq{defI} only by the presence of the Wilson 
term therefore one can again write it as
\begin{equation}
	 I_{W}[x; \{ \nhati \}]  = \sum _{h=1}^{2^{D}} I_{Wh}[x; \{ \nhati \}],
	\label{divideIW}
\end{equation}
where $I_{Wh}[x; \{ \nhati \}]$ is given by
\begin{eqnarray}	
	I_{Wh}[x; \{ \nhati \}] & = & e^{i\Kbarh . x}\int_{\frac{-\pi}{2a}}^{\frac{\pi}{2a}} 
	                     \frac{d^{D}k}{(2\pi)^{D}}\exp ( ik.x )
	                         \nonumber\\
	           && \times \exp \left( -i \frac{L}{N} \frac{\sin (\Kbarhmu a + \kmu 
	           a)}{a} \sumk \nhatikmu \nonumber \right.\\
	           && \left. - \sum_{\mu = 1}^{D} ( 1- \cos(\Kbarhmu 
	           a  +\kmu a) ) \right).
\label{defIWh}
\end{eqnarray}
In the limit $a \rightarrow 0$ one obtains
\begin{eqnarray}
	\limAzero I_{Wh}[x; \{ \nhati \}]  & = & e^{i\Kbarh . x} \exp( -\frac{L}{Da}
	                                   \sum_{\mu = 1}^{D} ( 1- \cos(\Kbarhmu 
	                                   a ) ) )
	\nonumber  \\
                                 	&  & \times \delta ^{D} (x - \frac{L}{N} \cos(\Kbarhmu a) \sumk 
	                                           \nhatikmu).
	\label{CuIWh}
\end{eqnarray}
From \Eq{CuIWh} one sees that in the limit $a \rightarrow 0$ only 
$I_{Wh}$ corresponding to $\Kbarh = 0$ will survive. Rest of the 
$I_{Wh}$ goes to zero, approximately as $\exp(-\frac{L}{a})$, and therefore 
their is no contribution from the ``doublers'' and one can write
\begin{equation}
	\limAzero  I_{W}[x; \{ \nhati \}] = \delta ^{D}(x - \frac{L}{N} \sumk \nhati)
	\label{CtnIW}.
\end{equation}
Substituting \Eq{CtnIW} in \Eq{FTGW} leads to the identification of $L$ as
the length of the path and $N$ as the number of steps of the random 
walk. Finally the propagator in the continuum limit can be written as
\begin{eqnarray}
	\limAzero \Delta_{W}(x, m) & = & \frac{a}{D} \sum_{N=1}^{\infty}
	                  \exp(-\frac{maN}{D})\Gtilde_{W} (N;a;x),
	\label{RWCtnW}  \\
	 \Gtilde_{W}(N; a; x)& = & \normFactor \sumdot \spinfactor 
	                 \nonumber\\
	                 &&\times \frac{\delta ^{D}(x-y, 
	                 a\sum_{k=1}^{N}\nhatik)}{a^{D}}.
	\label{RWGW}  
\end{eqnarray}

Thus, in the continuum limit one can represent the propagator of the 
Wilson fermions as a sum over paths  on a lattice, further more 
the representation is identical to the random walk representation of 
the continuum fermionic propagator. A path integral representation for 
the Wilson fermions in the form of a hopping-parameter expansion is well 
known~\cite{book2} and is 
similar to the path integral representation developed above. But 
conceptually the two differ, in obtaining \Eq{RWCtnW} no expansion in 
hopping parameter is done and the path integral representation emerges 
only in the continuum limit. Conceptually the path integral 
representation for the Wilson fermions is a way of transcribing 
the spin factor, present in the continuum path integral representation 
of a fermion, on to a lattice~\cite{PolyLH}.

Next I consider the propagator for the Kogut-Susskind fermions.
Since the development is almost identical to the 
previous cases only the final results will be stated. The propagator 
for the Kogut-Susskind fermions~\cite{book1, book2} can be written as
\begin{equation}
	\Delta_{KS}(p) = \frac{1}
	{m + (\gammamu \otimes {\bf 1})\frac{1}{b}\sin(\pmu b) +
	\frac{1}{b}\sum_{\mu=1}^{D}(1-\cos(\pmu b) ) (\gamma_{5} \otimes 
	t_{\mu}t_{5})},
	\label{KSprop}
\end{equation}
where $t_{\mu}$ are $2^{[\frac{D}{2}]}$ dimensional flavor matrices, 
and are given by
\begin{eqnarray}
	t_{\mu} & = & \gammamu ^{\dag},
	\nonumber \\
	t_{5}& = & \gamma_{5}^{\dag},
	\label{defTmat}
\end{eqnarray}
and $b = 2a$, $a$ being the lattice spacing. The propagator, \Eq{KSprop}, 
describes the propagation of $2^{[\frac{D}{2}]}$ degenerate flavors. 
Formally the Kogut-Susskind propagator differs from the Wilson 
propagator only by the presence of the flavor space 
matrices, and that the lattice spacing has been effectively doubled, 
so one can immediately write down the path integral representation for 
the Kogut-Susskind propagator as:
\begin{equation}
	\limAzero \Delta_{KS} (x, m) = \frac{b}{D}\sum_{N=1}^{\infty}
	                                                 \exp(-\frac{mbN}{D} ) 
	                                                 \Gtilde_{KS} (N; b; 
	                                                 x)
	\label{RWKS}
\end{equation}
where $\Gtilde_{KS}$ is given by
\begin{eqnarray}
	\Gtilde_{KS} (N; b; x) & = & \normFactor \sumdot ( \spinfactor ) \otimes {\bf 
	1}
	\nonumber  \\
	 &  & \times \frac{\delta^{D} (x, b\sumk \nhatik )}{b^{D}}.
	\label{GKS}
\end{eqnarray}
In \Eq{RWKS} one has the path integral representation for the propagator 
of $2^{[\frac{D}{2}]}$ degenerate fermions.  The paths are defined 
on a blocked lattice with a step size of $b = 2a$. 

The path integral representation of the Kogut-Susskind propagator is, 
apart from the fact that it represents the propagation of 
$2^{[\frac{D}{2}]}$ degenerate fermions, identical to the path integral 
representation of the continuum fermions. The fact that the path 
integral representation of both the Wilson and the Kogut-Susskind fermions,
which are derived from different lattice actions and have different 
lattice symmetries, 
coincides with each other is not surprising as 
these representations appears only 
in the continuum limit. What is, perhaps, surprising is the fact that 
the continuum fermionic propagator can be represented as a sum over 
paths even when the paths are restricted to a lattice.

\section{Path Integral Representation of Green's \\
 Function in QCD and the Quark Model}

As stated in the introduction one of the motivations for developing a path 
integral representation for the lattice fermionic propagators was to 
look for a formalism that can provide some heuristic insight into 
the role of quark degrees of freedom in QCD. To this end, in this section, 
I will write the Green's functions of lattice QCD involving quark degrees 
of freedom as a sum over the paths of valence quarks thus making a 
connection with the heuristic language of quark model.

Consider a meson propagator build from  $\bar{\psi}(x)\Gamma\psi (x)$ as the 
interpolating field for a quark-antiquark meson with $\Gamma$ determining 
the spin, flavor, and the parity of the meson~\cite{book1, book2}. 
Such a propagator can be written as
\begin{eqnarray}
	\langle \bar{\psi}(x)\Gamma\psi (x)\bar{\psi}(y)\Gamma\psi (y)\rangle
	&=& Z^{-1}\int DU \exp(-S_{eff}(U) ) \nonumber \\
	 &&\times ( \Tr[\Delta (x, y, U) \Gamma\Delta (y, x, U) \Gamma]
	\nonumber \\
	&&- \Tr[\Delta(x, x, U) \Gamma] \Tr[\Delta (y, y, U) \Gamma] )
	\label{solveFermions}
\end{eqnarray}
where the quark degrees of freedom have been integrated out giving 
rise to an effective gauge action and fermionic propagators.
The partition function, 
$Z$, and  the effective gauge action, $S_{eff}(U)$, being  defined by
\begin{eqnarray}
	Z & = & \int DU \exp(-S_{eff} (U) ),
	\label{Z}  \\
	S_{eff} (U) & = & S_{g} (U) + \ln {\rm Det} \Delta ^{-1} (U).
	\label{Seff}
\end{eqnarray}
$S_{g} (U)$  represents the lattice regularized action for the gauge 
field $U$~\cite{book1, book2}. For the sake of definiteness, the fermionic determinant is 
assumed to be defined for the Wilson fermions.
In what follows it will be convenient to use the following notation 
to represent the functional integration over the gauge fields
\begin{equation}
	\langle f(U) \rangle_{U} = Z^{-1}\int DU \exp(-S_{eff}(U)) f(U) .
	\label{Uintegral}
\end{equation}
Also the path integral representation of the quark propagator, 
\Eq{CferLatG}, will 
be written in the following compact notation
\begin{eqnarray}
	\Delta (x, y, U) & = & \sum_{l_{xy}} \exp (-\mu l )  \Phi(l_{xy}) 
	U(l_{xy}),
	\label{myProp}  
\end{eqnarray}
where $l_{xy}$ denotes a path on a lattice of length $l$, in lattice 
units, starting at $y$ and ending at $x$, $\mu$ is a measure of the bare 
mass and is given by
\begin{equation}
	\mu = \frac{ma}{D},
	\label{bareMass}
\end{equation}
while the spin factor $\Phi(l_{xy})$ and the gauge field factor 
$U(l_{xy})$ are 
given explicitly by \Eq{LGGtilde}.

In the above defined notations \Eq{solveFermions} can be written as
\begin{eqnarray}
	 \langle \bar{\psi}(x)\Gamma\psi (x)\bar{\psi}(y)\Gamma\psi (y)\rangle
	& = & \langle \Tr[\Delta (x, y, U) \Gamma \Delta (y, x, U) \Gamma] \rangle_{U}
	\nonumber  \\
	&   & - \langle \Tr[\Delta(x, x, U) \Gamma] \Tr[\Delta (y, y, U) 
	\Gamma] \rangle_{U}.
	\label{mesonProp}
\end{eqnarray}
Consider the two terms appearing on the right hand side of 
\Eq{mesonProp} separately and replace the fermionic propagators 
appearing in them by their path integral representation~\Eq{myProp}. 
The first term can be written as
\begin{eqnarray}
	 \langle \Tr[\Delta (x, y, U) \Gamma \Delta (y, x, U) \Gamma] \rangle_{U}& = & 
	 \sum_{l_{xy}, l_{yx}} \exp (-\mu (l_{xy} + l_{yx} ) )\nonumber \\
	&  &\times \Tr[\Phi(l_{xy})\Gamma\Phi(l_{yx})\Gamma] \nonumber \\
	&  & \times 
	        \langle \Tr ( U(l_{xy}) U(l_{yx}) )\rangle_{U}
	\label{conMesonProp}.
\end{eqnarray}
Similarly the second term can be written as 
\begin{eqnarray}
	\langle \Tr[\Delta(x, x, U) \Gamma] \Tr[\Delta (y, y, U) 
	\Gamma] \rangle_{U}. & = & \sum_{l_{xx}, l_{yy}} \exp(-\mu 
	(l_{xx} + l_{yy})\nonumber \\
	 &  & \times \Tr[\Phi(l_{xx})\Gamma] \Tr[\Phi(l_{yy})\Gamma]
	\nonumber  \\
	 &  & \times \langle \Tr U(l_{xx}) \Tr U(l_{yy}) \rangle_{U}.
	\label{disconMesonProp}
\end{eqnarray}

Another quantity of interest is the order parameter for the chiral symmetry, 
$\bar{\psi} \psi$, which can be written as a sum over closed paths
 \begin{eqnarray}
 	\langle \bar{\psi}(0) \psi(0) \rangle & = & \langle 
 	\Tr[\Delta(0,0,U)] \rangle_{U}
 	\nonumber  \\
 	 & = & \sum_{l_{00}}\exp(-\mu l_{00} ) \Tr[\Phi(l_{00})] \langle 
 	 \Tr U(l_{00})\rangle_{U}.
 	\label{psibarpsi}
 \end{eqnarray}
 
In a similar manner all the Green's functions involving quark 
fields can be written as a sum over one or more closed (valence) quark paths, each path 
being weighted by its length, its spin factor, and the expectation 
value of the path ordered product of the gauge fields along the quark 
paths. The effect of the dynamical or the sea quarks being included in 
the expectation value of the gauge field factor. 
The advantage of writing the quark Green's function using the path 
integral representation of the quark propagator is that it separates 
the contribution coming from the motion of the valence quark, in a 
sense the kinematics, from the contribution coming from the 
fluctuations in the gauge fields and the quark degrees of freedom.
As a possible application of the random walk representation 
of the fermionic propagators, consider the relationship between QCD and the 
constituent quark model~\cite{Georgi}.
In the chiral limit the constituent quark picture for pion is not 
useful, pion being the Nambu-Goldstone boson, while it is a 
good approximation for the rho meson.  One would like to understand, starting 
from QCD, the emergence of the constituent quark picture for the rho 
and its absence for the pion. 
For this purpose consider the connected part of the meson propagator
\Eq{conMesonProp}, {\it a priori }
all possible paths contribute to this propagator but one expects 
that because of the confinement the paths in which quark and 
anti-quark are widely separated are strongly suppressed. This 
motivates us to restrict the sum in \Eq{conMesonProp} to a 
class of paths in which the quark and the anti-quark are separated 
at most by, say $r_{max}$, where $r_{max}$ is some measure of the 
meson size. One expects that the constituent quark model should emerge 
at a coarser level of description~\cite{PHWilson90} and such a coarser 
description should appear as a result of a renormalization group 
transformation. In the present context one could try and see if the 
constituent quark model emerges after the paths have been blocked, where 
blocking of paths is in analogy with the blocking of spins under 
the real space renormalization group transformation~\cite{Wilson79}. 
One possible way of blocking the paths is by summing over the paths 
in the neighborhood of a given path belonging to the above defined 
class. In particular consider a path $l_{xyx}$ (where 
$l_{xyx}$ being a compact notation for the closed path 
$l_{xy}.l_{yx}$), imagine that 
that this path is enclosed in a tube of radius $r_{b}\approx 2 
r_{max}$, and then sum over all the paths which lie inside this tube to 
create a blocked path. Denoting the paths lying inside the tube by 
$\tilde l$, one can \textit{conjecture} that in the case of the rho meson the 
sum over these paths,
\begin{equation}
	\sum_{\tilde l_{xyx}}\exp(-\mu \tilde l)
	                             \Tr [\Phi(\tilde l_{xy}) \Gamma_{\rho} 
         	                             \Phi(\tilde l_{yx}) \Gamma_{\rho}]
         	                             \langle \Tr U(\tilde l_{xyx}) \rangle_{U} ,
	\label{rho}	
\end{equation}
can be approximated by a single path with the following weight factors
\begin{equation}
	\exp (-m_{eff} (l_{xy}+l_{yx}) \Tr 
	[\Phi(l_{xy})\Gamma_{\rho}\Phi(l_{yx})\Gamma_{\rho}] \exp(-V_{eff} 
	(l_{xy}, l_{yx}))
	\label{rhoQuarkModel}
\end{equation}
where $m_{eff}$ is a measure of the constituent quark mass and $\exp(-V_{eff})$
 is the weight factor coming from the confining 
potential of the constituent quark model. The simplest possible guess 
for $V_{eff}$ would be $\sigma A_{xyx}$ where $\sigma$ is the string 
tension and $A_{xyx}$ is the minimal area enclosed by the closed path 
$l_{xyx}$.

On the other hand for pions one 
would conjecture that the blocked paths should be approximated by a single 
effective path with the following weight factors 
\begin{math}
	\exp(-m_{\pi}l_{xy}),
\end{math}
where $m_{\pi}$ being the measure of the pion mass on the lattice
which in turn should be related to the expectation value of the order 
parameter for the chiral symmetry breaking, $\langle \bar \psi \psi \rangle$.
Such heuristic conjectures can be numerically tested, for the sum in 
\Eq{rho} is well defined on a lattice and is a small subset of all 
possible paths connecting point $x$ and $y$. Perhaps the biggest obstacle 
in such a numerical 
investigation is the requirement of large lattice sizes, for the random 
walk representation appears only as one approaches the continuum limit. For finite 
lattices one will have to understand how to separate the lattice 
artifacts from the continuum physics. A first step in such a 
direction would be to use the above formalism for exploring 
the dynamic of lattice QCD in two dimensions.

\section{Conclusions}

Path integral or the random walk representation for the fermionic 
propagator was developed with two motivations in mind. One to try and 
understand the phenomena of fermion doubling on a lattice in the 
language of sum over paths, and other to look for a formalism for the 
fermionic degrees of freedom that might be more susceptible to 
heuristic investigations.

In the first respect, it was shown that a naive lattice fermionic 
propagator can be written as a sum over paths and with each path there 
are associated not one but $2^{D}$ spin factors in $D$ dimensions. 
As a result the naive propagator represents the propagation of 
$2^{D}$ fermions. This is in contrast to the case of the continuum 
fermionic propagator which can be represented as a sum over paths on 
a lattice with a unique spin factor associated with each path. In 
the case of the Wilson and the Kogut-Susskind fermions the path 
integral representation emerges in the continuum limit and coincides 
with the path integral representation of the continuum fermionic 
propagator.  In this manner the path integral representation provides an additional 
insight into the relationship between various lattice fermionic 
propagators and their continuum counterpart, this may be of some use in 
exploring new ways of representing fermionic degrees of freedom on a 
lattice.

As to the second motivation, it was shown that using path integral 
representation for the quark propagators, the Green's functions of 
QCD can be written as a sum over the paths of the valence quarks. Such an 
representation of the Green's functions allows for the possibility of 
delineating the important paths that contribute to a Green's function. 
Also it allows for the possibility of implementing the ideas of  
real space renormalization group on variables which are easy to 
visualize and are close to the heuristic description of QCD, 
namely the paths of the valence quarks.  This was illustrated by showing how one 
can use numerical simulations to understand the relationship between 
QCD and the constituent quark model.

The path integral representation of the lattice fermionic propagator 
is unlikely to be useful for an accurate numerical solution of lattice QCD 
but it can be a useful tool for testing 
heuristic insights and for developing new intuitions. 
\section*{Acknowledgments}

I would like to thank the members of the Physics Department, University 
of Utah, where this work was initiated. At the University of Utah 
I am particularly indebted to Carlton DeTar 
for his support and encouragement when they were needed most.

\pagebreak

\end{document}